\documentclass{article}[12pt]
\usepackage{amsmath,amssymb}
\usepackage{graphicx}%
\usepackage{xcolor}
\usepackage{verbatim}
\usepackage{multirow}

\newtheorem{stat}{Утверждение}
\newtheorem{property}{Свойство}

\newcommand{\TwoFigss}[4]{%
\begin{flushleft}
\begin{tabular}{lr}
\parbox{7cm}{\centerline{\includegraphics[width=7cm]{#1}}}  & \parbox{7cm}{\centerline{\includegraphics[width=7cm]{#2}}}  \\
\parbox{7cm}{\vspace{7pt}\refstepcounter{figure}Fig. \thefigure.\quad #3\vfill} & \parbox{7cm}{\vspace{7pt}\refstepcounter{figure}Fig. \thefigure.\quad #4\vfill}\\
\end{tabular}
\end{flushleft}
\vspace{8pt}
}
\newcounter{strochka}

\newcounter{spisok}
\begin{document}

\begin{center}
{\bf \Large Yu. G  Ignat'ev\footnote{Institute of Physics, Kazan Federal University, Research Laboratory of Cosmology, 420008 Russia, Kazan, st. Kremlevskaya 18; email: yurii.ignatev.1947@yandex.ru} }\\[12pt]
{\bf \Large Self-gravitating Higgs field of scalar charge. II. Asymmetric scalar doublet.} \\[12pt]
\end{center}

\abstract{The self-gravitating Higgs field of a scalar charge is studied for the case of an asymmetric scalar doublet containing, along with a canonical and a phantom component. It is shown that in the zero and first approximation of the smallness of the canonical and phantom scalar charges, the gravitational field of the scalar charge is described by the Schwarzschild-de Sitter metric with a cosmological constant determined by the stable equilibrium point - the vacuum potential of the canonical Higgs field and the zero value of the scalar potential. An equation for the perturbation of the stable value of the potential is obtained and studied, and the asymptotic behavior in the near and far zones is found. The averaging of microscopic oscillations of the scalar field is carried out and it is shown that the sign of the contribution of microscopic oscillations to the macroscopic energy of the scalar field is completely determined by the values of the fundamental constants of the Higgs potential of the asymmetric scalar doublet. Particular attention is paid to the case when the contribution of oscillations to the macroscopic energy and pressure densities is strictly equal to zero. Possible applications of the obtained solutions are discussed.\\

{\bf Keywords}: scalarly charged black hole, asymmetric scalar doublet, scalar Higgs field, asymptotic behavior, macroscopic characteristics, doublet orientation.
}


%
\section{Introduction}
In \cite{Yu_TMF24} the self-gravitating Higgs field of a single massive scalar charge was studied. In this work, it was shown that in the first approximation of the smallness of the scalar charge, its scalar field is close to the stable equilibrium point $\Phi=\pm m_s/\sqrt{\alpha}$, and the gravitational field is described by the Schwarzschild - de Sitter metric. In this case, microscopic oscillations with negative average energy arise in the far zone $r\to \infty$. This issue requires a deeper and broader study, and therefore in this work we study a similar problem for the asymmetric scalar doublet \cite{YuKokh_TMF}.

The Lagrange function $L$ of an asymmetric scalar Higgs doublet \emph{with non-interacting canonical $\Phi$ and phantom $\varphi$ components} is (see, for example, \cite{YuKokh_TMF} and the bibliography contained there)\footnote{Here and then Latin letters run through the values $\overline{1,4}$, Greek letters - $\overline{1,3}$. The Planck system of units $G=c=\hbar=1$ is used throughout.
\label{Plank_units}}:
\begin{eqnarray} \label{L_s}
L\equiv L_c+L_f=L_s=\frac{1}{16\pi}(g^{ik} \Phi_{,i} \Phi_{,k} -2V(\Phi))+\frac{1}{16\pi}(-g^{ik} \varphi_{,i} \varphi_{,k} -2\upsilon(\varphi)),
\end{eqnarray}
where
\begin{eqnarray}
\label{Higgs}
V(\Phi)=-\frac{\alpha}{4} \left(\Phi^{2} -\frac{m_s^{2}}{\alpha}\right)^{2}; & \displaystyle \upsilon(\varphi)=-\frac{\beta}{4} \left(\varphi^{2} -\frac{\mu_s^{2}}{\beta}\right)^{2}
\end{eqnarray}
are the potential energies of the canonical and phantom scalar fields, respectively, $\alpha,\beta$ are their self-interaction constants, $m_s,\mu_s$ are the boson masses.
The energy tensor - momentum of the scalar doublet with respect to the Lagrange function \eqref{L_s} is:
\begin{eqnarray}\label{T_s}
T^i_{k} \equiv  T^i_{k (c)}+T^i_{k (f)}=\frac{1}{16\pi}\bigl(2\Phi^{,i}\Phi_{,k}- \delta^i_k\Phi_{,j} \Phi^{,j}+2V(\Phi)\delta^i_k \bigr)+
\frac{1}{16\pi}\bigl(-2\varphi^{,i}\varphi_{,k}+ \delta^i_k\varphi_{,j} \varphi^{,j}+2\upsilon(\varphi)\delta^i_k \bigr).
\end{eqnarray}

Further, Einstein's equations take the form:
\begin{equation}\label{Eq_Einst_G}
R^i_k-\frac{1}{2}\delta^i_k R=8\pi T^i_k + \delta^i_k \Lambda_0,
\end{equation}
where $\Lambda_0$ is the initial value of the cosmological constant, associated with its observed value $\Lambda$, obtained by removing the constant terms in the potential energy, by the relation:
\begin{equation}\label{lambda0->Lambda}
\Lambda=\Lambda_0-\frac{1}{4}\frac{m_s^4}{\alpha}-\frac{1}{4}\frac{\mu_s^4}{\beta}.
\end{equation}

In curvature coordinates (see, for example, \cite{Land_Field})
\begin{equation}\label{metric_stat}
ds^2=\mathrm{e}^{\nu(r)}dt^2-\mathrm{e}^{\lambda(r)}dr^2-r^2 d\Omega^.
\end{equation}
\begin{eqnarray}\label{T^i_k}
T^1_1=-\frac{\mathrm{e}^{-\lambda(r)}}{16\pi}({\Phi'}^2-{\varphi'}^2)-\frac{\alpha}{32\pi}\left(\Phi^2-\frac{m^2_s}{\alpha}\right)^2-\frac{\beta}{32\pi}\left(\varphi^2-\frac{\mu^2_s}{\beta}\right)^2\quad (\equiv -p_\parallel),\nonumber\\
T^2_2=T^3_3=T^4_4=\frac{\mathrm{e}^{-\lambda(r)}}{16\pi}({\Phi'}^2-{\varphi'}^2)-\frac{\alpha}{32\pi}\left(\Phi^2-\frac{m^2_s}{\alpha}\right)^2-\frac{\beta}{32\pi}\left(\varphi^2-\frac{\mu^2_s}{\beta}\right)^2\quad (\equiv -p_\perp=\varepsilon),
\end{eqnarray}
where $p_\parallel$ is the radial pressure, $p_\perp$ is the pressure along the surface of the sphere, $\varepsilon$ is the energy density of the scalar field.

\section{Spherically symmetric scalar field with the Higgs potential of a point binary scalar charge in the pseudo-Euclidean metric}

Let us now study the gravitational field generated by a binary scalar charge with the Higgs potential in the metric \eqref{metric_stat}. The equations of the scalar canonical and phantom Higgs field $\Phi(r)$ in this metric have the form:
\begin{eqnarray}\label{Eq_C}
\displaystyle \frac{1}{r^2}\frac{d}{dr}\left(r^2\mathrm{e}^{\frac{\nu-\lambda}{2}}\frac{d}{dr}\Phi\right)-\mathrm{e}^{\frac{\nu-\lambda}{2}}\Phi(m^2_s-\alpha\Phi^2)=0;\\
\label{Eq_F}
\displaystyle \frac{1}{r^2}\frac{d}{dr}\left(r^2\mathrm{e}^{\frac{\nu-\lambda}{2}}\frac{d}{dr}\varphi\right)+\mathrm{e}^{\frac{\nu-\lambda}{2}}\varphi(\mu^2_s-\beta\varphi^2)=0.
\end{eqnarray}
\subsection{Stability of one-dimensional solutions}
Before we begin to find spherically symmetric solutions to field equations, we examine the stability of one-dimensional solutions to the corresponding field equations. Wave equations for one-dimensional scalar fields in the pseudo-Euclidean metric have the form:
\begin{eqnarray}\label{Eq_C0}
\frac{\partial^2}{{\partial t}^2}\Phi-\frac{\partial^2}{{\partial x}^2}\Phi+\Phi(m^2_s-\alpha\Phi^2)=0;\\
\label{Eq_F0}
-\frac{\partial^2}{{\partial t}^2}\varphi+\frac{\partial^2}{{\partial x}^2}\varphi+\varphi(\mu^2_s-\beta\varphi^2)=0.
\end{eqnarray}
To apply the qualitative theory of ordinary differential equations to our problem, we consider two problems separately; in the first of them, the scalar fields depend only on the time variable $t$ - $\Phi(t)$, $\varphi(t)$.
Let's call this situation a \emph{T-situation}. In the second problem, the scalar fields depend only on the spatial coordinate $x$ -- $\Phi(x)$, $\varphi(x)$. Let's call this situation an \emph{R-situation}.
\subsubsection{T-situation}
In this case, which corresponds, for example, to the cosmological situation, the system of equations \eqref{Eq_C0} -- \eqref{Eq_F0} takes the form:
\begin{eqnarray}\label{Eq_C0T}
\ddot{\Phi}+\Phi(m^2_s-\alpha\Phi^2)=0 & \Rightarrow \dot{\Phi}=Z; & \dot{Z}=-\Phi(m^2_s-\alpha\Phi^2);\\
\label{Eq_F0T}
-\ddot{\varphi}+\varphi(\mu^2_s-\beta\varphi^2)=0 & \Rightarrow \dot{\varphi}=z; & \dot{z}=\varphi(\mu^2_s-\beta\varphi^2).
\end{eqnarray}
The main matrices of dynamic systems (see, for example, \cite{Bogoyav}) \eqref{Eq_C0T} -- \eqref{Eq_F0T} have the form:
\begin{eqnarray}\label{A0}
\mathbf{A}_C=\left(\begin{array}{cc}
0 & 1 \\
-m^2+3\alpha\Phi^2 & 0
\end{array}
\right); & \displaystyle
\mathbf{A}_F=\left(\begin{array}{cc}
0 & 1 \\
\mu^2-3\beta\varphi^2 & 0
\end{array}
\right).
\end{eqnarray}
The eigenvalues of these matrices are equal;
\begin{equation}\label{lambda_i}
\lambda^{(c)}_{1,2}=\pm i\sqrt{m^2-3\alpha\Phi^2}; \quad \lambda^{(f)}_{1,2}=\pm \sqrt{\mu^2-3\beta\varphi^2},
\end{equation}
and the coordinates of the singular points $M^{(c)}(\Phi_0,Z_0)$ and $M^{(f)}(\varphi_0,z_0)$ ($\dot{\Phi}=0, \dot{Z }=0,\ \dot{\varphi}=0, \dot{z}=0$) are equal
\begin{eqnarray}\label{sing_points}
M^{(c)}_0=[0,0],\; M^{(c)}_\pm= \left[\pm\frac{m^s}{\sqrt{\alpha}},0\right]; & \displaystyle M^{(f)}_0=[0,0],\; M^{(f)}_\pm= \left[\pm\frac{\mu^s}{\sqrt{\beta}},0\right].
\end{eqnarray}
Thus, we obtain the values of the eigenvalues \eqref{lambda_i} at singular points of the system:
\begin{eqnarray}\label{lambda_Tc}
M^{(c)}_0(0,0): \lambda=\pm i m_s; & \displaystyle M^{(c)}_\pm\left(\pm\frac{m_s}{\sqrt{\alpha}},0\right): \lambda=\pm\sqrt{2}m_s;\\
\label{lambda_Tf}
M^{(f)}_0(0,0): \lambda=\pm \mu_s; & \displaystyle M^{(f)}_\pm\left(\pm\frac{\mu_s}{\sqrt{\beta}},0\right): \lambda=\pm i\sqrt{2}\mu_s
\end{eqnarray}

Thus, in the T-situation for a canonical scalar field, the zero point $M^{(c)}_0(0,0)$ is a stable center, and the points $M^{(c)}_\pm$ are saddle points, i.e. That is, they correspond to unstable solutions. For a phantom field, the situation is the opposite: the zero point $M^{(f)}_0(0,0)$ is a saddle (unstable), and the points $M^{(f)}_\pm$ are stable centers. In such a system, over time the solution should tend to
to sustainable, i.e.,
\begin{eqnarray}\label{t8}
\mathbf{T}:\;  t\to+\infty \Rightarrow & \Phi\to 0; & \varphi\to \varphi_\pm=\pm\frac{\mu_s}{\sqrt{\beta}};
\end{eqnarray}
\subsubsection{R-situation}
In this case, which corresponds, for example, to an astrophysical situation, the system of equations \eqref{Eq_C0} -- \eqref{Eq_F0} takes the form:
\begin{eqnarray}\label{Eq_C0R}
-\Phi''+\Phi(m^2_s-\alpha\Phi^2)=0 & \Rightarrow \Phi'=Z; & Z'=\Phi(m^2_s-\alpha\Phi^2);\\
\label{Eq_F0R}
\varphi''+\varphi(\mu^2_s-\beta\varphi^2)=0 & \Rightarrow \varphi'=z; & z'=-\varphi(\mu^2_s-\beta\varphi^2).
\end{eqnarray}
In this case, the coordinates of the singular points will not change, i.e., they will coincide with \eqref{sing_points}, however, their character will change to the opposite in comparison with \eqref{lambda_Tc} -- \eqref{lambda_Tf}:
\begin{eqnarray}\label{lambda_Rc}
M^{(c)}_0(0,0): \lambda=\pm m_s; & \displaystyle M^{(c)}_\pm\left(\pm\frac{m_s}{\sqrt{\alpha}},0\right): \lambda=\pm i\sqrt{2}m_s;\\
\label{lambda_Rf}
M^{(f)}_0(0,0): \lambda=\pm i\mu_s; & \displaystyle M^{(f)}_\pm\left(\pm\frac{\mu_s}{\sqrt{\beta}},0\right): \lambda=\pm \sqrt{2}\mu_s.
\end{eqnarray}

Thus, in the R-situation for a canonical scalar field, the zero point $M^{(c)}_0(0,0)$ is a saddle point, i.e., unstable, and the points $M^{(c)}_\pm$ are centers, i.e., they correspond to stable periodic solutions. For a phantom field, the situation is the opposite: the zero point $M^{(f)}_0(0,0)$ is a stable center, and the points $M^{(f)}_\pm$ are saddle points, i.e., unstable . In such a system, over time, the solution should tend to be stable, i.e.
\begin{eqnarray}\label{x8}
\mathbf{R}:\;  x\to+\infty \Rightarrow & \displaystyle \Phi\to \Phi_\pm=\pm\frac{m_s}{\sqrt{\alpha}}; & \varphi\to 0.
\end{eqnarray}

In this work, we are interested in the R-situation, so we are obliged to select solutions with asymptotic behavior \eqref{x8}. Solutions that do not have such asymptotic behavior will be unstable, i.e., with small changes in the initial conditions of the Cauchy problem, these solutions can infinitely differ from each other at spatial infinity.

\subsubsection{Dispersion relations}
Let us now consider the general behavior of solutions $\{\Phi(x,t),\varphi(x,t)\}$ of the equations \eqref{Eq_C0} -- \eqref{Eq_F0} near singular points \eqref{sing_points} of one-dimensional systems \eqref{Eq_C0T} -- \eqref{Eq_F0T} and \eqref{Eq_C0R} -- \eqref{Eq_F0R}, presenting solutions in the form
\begin{eqnarray}\label{i_omega_t}
\Phi=\Phi_0+S \mathrm{e}^{i\omega t+ikx}; & \varphi=\varphi_0+s \mathrm{e}^{i\omega t+ikx}, & (S,s\ll 1),
\end{eqnarray}
where $(\Phi_0,0)$ and $(\varphi_0,0)$ are the coordinates of singular points. Substituting \eqref{i_omega_t} into the equations of the equations \eqref{Eq_C0} -- \eqref{Eq_F0} and linearizing their smallnesses $S,s$, we obtain a system of homogeneous algebraic equations for the amplitudes $S,c$, the condition for nontrivial solvability of which is: called dispersion relations, which in our case will have the following form:
\begin{eqnarray}
\label{disp_F=F0}
\Phi_0=\pm \frac{m_s}{\sqrt{\alpha}}: & S(-\omega^2+k^2-2m_s^2)=0\Rightarrow \omega=\pm \sqrt{k^2-2m_s^2}; & S=S_0 \exp(\pm i\sqrt{k^2-2m_s^2}t + ikx);\\
\label{disp_F=0}
\Phi_0=0: & S(-\omega^2+k^2+m_s^2)=0\Rightarrow \omega=\pm \sqrt{k^2+m_s^2} & S=S_0 \exp(\pm i\sqrt{k^2+m_s^2}t\pm ikx);\\
\label{disp_f=f0}
\varphi0=\pm \frac{\mu_s}{\sqrt{\beta}}:  & s(\omega^2-k^2-2\mu_s^2)=0; \Rightarrow \omega=\pm \sqrt{k^2+2\mu_s^2};   & s=s_0 \exp(\pm i\sqrt{k^2+2m_s^2}t + ikx);\\
\label{disp_f=0}
\varphi0=0:  & s(\omega^2-k^2+\mu_s^2)=0; \Rightarrow \omega=\pm \sqrt{k^2-\mu_s^2};   & s=s_0 \exp(\pm i \sqrt{k^2-\mu_s^2}t + ikx).
\end{eqnarray}
Thus, solutions with $\Phi_0=0$ and $\varphi_0=\pm \mu_s/\sqrt{\beta}$ are pairs of retarded and advanced waves, while solutions with $\Phi_0=\pm m_s/ \sqrt{\alpha}$ and $\varphi_0=0$ in areas
$|k|<\sqrt{2}m_s$ and $|k|<\mu_s$ are standing waves with an amplitude infinitely increasing with time, i.e., these solutions are unstable in the long-wave sector of disturbances. So, the criteria for the stability of solutions in the general case are similar to the criteria for the stability of solutions in the T-situation. Note that the long-wave sector of perturbations precisely describes the behavior of solutions at infinity.
\subsubsection{Centrally symmetric solutions}
Let's study static centrally symmetric solutions of field equations. In the pseudo-Euclidean metric, the equations \eqref{Eq_C} -- \eqref{Eq_F} in the presence of a singular binary scalar charge ${Q,q}$ take the form:
\begin{eqnarray}\label{Eq_Cr0}
\frac{1}{r^2}\frac{d}{dr}\biggl(r^2\frac{d\Phi}{dr}\biggr)-m^2\Phi+\alpha\Phi^3=-8\pi \mathrm{Q}\delta(\mathbf{r});\\
\label{Eq_Fr0}
\frac{1}{r^2}\frac{d}{dr}\biggl(r^2\frac{d\varphi}{dr}\biggr)+\mu^2\varphi-\beta\varphi^3=8\pi \mathrm{q}\delta(\mathbf{r}).
\end{eqnarray}
It is necessary to keep in mind the following ratio:
\begin{equation}\label{1/r}
\frac{1}{r^2}\frac{d}{dr}\biggl(r^2\frac{d}{dr}\frac{1}{r}\biggr)=-4\pi\delta(\mathbf{r}).
\end{equation}

According to the above, we will look for solutions to the equations \eqref{Eq_Cr0} -- \eqref{Eq_Fr0} that are close to stable at infinity \eqref{x8}, assuming
\begin{equation}\label{F,f=F0,f0+e}
\Phi=\Phi_\pm+\phi,\; \phi\ll 1; \qquad \varphi\ll 1.
\end{equation}
Expanding the equations \eqref{Eq_Cr0} -- \eqref{Eq_Fr0} in small $\phi(r),\varphi(r)$, we obtain in the linear approximation the equations
\begin{eqnarray}\label{Eq01_Cr}
\frac{1}{r^2}\frac{d}{dr}\biggl(r^2\frac{d\phi}{dr}\biggr)+2m^2_s\phi=-8\pi \mathrm{Q}\delta(\mathbf{r});\\
\label{Eq01_Fr}
\frac{1}{r^2}\frac{d}{dr}\biggl(r^2\frac{d\varphi}{dr}\biggr)+\mu^2_s\varphi=8\pi \mathrm{q}\delta(\mathbf{r}).
\end{eqnarray}

Solving the equations \eqref{Eq01_Cr} -- \eqref{Eq01_Fr} taking into account the relation \eqref{1/r}, we will finally find
\begin{eqnarray}\label{Phi,varphi=0+e}
\Phi\backsimeq \pm \frac{m_s}{\sqrt{\alpha}}+\frac{2Q}{r}\cos(\sqrt{2}m_s r); & \displaystyle \varphi\backsimeq -\frac{2q}{r}\cos(\mu_s r).
\end{eqnarray}
The derivatives of the potentials near the charge are equal
\begin{equation}\label{phi,varphi_R->0}
\left.\frac{d\Phi}{dr}\right|_{r\to0}\backsimeq -\frac{2Q}{r^2}\quad (\sqrt{2}m_s r\ll1); \quad \left.\frac{d\varphi}{dr}\right|_{r\to0}\backsimeq \frac{2q}{r^2}, \quad (\mu_s r\ll1),
\end{equation}
i.e., they behave like the electric field strength of a point charge.

The presence of a \emph{fundamental} scalar field with the Higgs potential fundamentally changes the physical picture. Now the vacuum state corresponds to one of the stable points of the Higgs potential
\begin{equation}\label{stab_F,f}
\Phi=\Phi_0=\pm\frac{m_s}{\sqrt{\alpha}}, \quad \varphi=\varphi_0=0.
\end{equation}
Calculating according to \eqref{T^i_k} the energy density of the scalar doublet at this point, we find:
\begin{equation}\label{stabilR}
\varepsilon_0=-\frac{\mu^4_s}{32\pi\beta}.
\end{equation}
According to the formula for renormalizing the cosmological constant \eqref{lambda0->Lambda} it is this value that corresponds to the contribution of the phantom field to the renormalized cosmological constant at a stable point of the system
\begin{equation}\label{lambda0->Lambda_1}
\delta\Lambda=8\pi\varepsilon_0=-\frac{1}{4}\frac{\mu_s^4}{\beta}.
\end{equation}

Note that in the work \cite{Scalar_Charge} the field equations for single scalar charges with the Higgs potential in pseudo-Euclidean space were integrated using numerical methods. In this work, it was shown that when initial conditions are specified far from the stable point \eqref{stab_F,f}, the solution exponentially quickly tends to this point, and the transition process is accompanied by oscillations that decay near the stable point.

\section{Self-gravitating asymmetric scalar Higgs doublet}
\subsection{Field equations}
Taking into account the above, we study the solution to the complete problem of a self-gravitating scalar Higgs field. Nontrivial combinations of Einstein's equations with a cosmological constant in the metric \eqref{metric_stat}\footnote{These are combinations of the equations $^1_1$, $^4_4$ and the scalar field equation.} can be reduced to the form:
\begin{eqnarray}\label{Eq_A}
r(\Phi'^2-\varphi'^2)+(\lambda+\nu)'=0;\\
\label{Eq_B}
\mathrm{e}^\lambda-1-\frac{r}{2}(\lambda'-\nu')-\!r^2\mathrm{e}^\lambda \left[\Lambda_0+\frac{\alpha}{4}\left(\Phi^2-\frac{m^2_s}{\alpha}\right)^2
+\frac{\beta}{4}\left(\varphi^2-\frac{\mu^2_s}{\beta}\right)^2\right]=0.
\end{eqnarray}

We will look for solutions to the system of equations \eqref{Eq_C}, \eqref{Eq_A}, \eqref{Eq_B} that are close to stable \eqref{stabilR}, assuming
\begin{eqnarray}\label{F,f->stabil}
\Phi(r)=\Phi_0+\phi(r)\equiv \pm \frac{m_s}{\sqrt{\alpha}}+\phi(r), \quad \phi(r)\ll1; & \varphi(r)\ll 1.
\end{eqnarray}
Then, in the zero approximation, by the smallness of $\phi(r),\varphi(r)$, the field equations for scalar fields \eqref{Eq_Cr0} -- \eqref{Eq_Fr0} become identical, and the equation \eqref{Eq_A} gives
\begin{equation}\label{lambda=-nu}
\lambda=-\nu.
\end{equation}
As a result, the equation \eqref{Eq_B} will be reduced to a closed-loop equation for $\nu$ (or $\lambda$)
\begin{eqnarray}
\label{Eq_B0}
r\nu'+1+\mathrm{e}^{-\nu}(1-\tilde{\Lambda} r^2)=0,
\end{eqnarray}
where \emph{in direct accordance with \eqref{lambda0->Lambda_1}}
\begin{equation}\label{Lambda_tilde}
\tilde{\Lambda}\equiv \Lambda_0+\frac{\mu_s^4}{4\beta}.
\end{equation}
Solving the equation \eqref{Eq_B0} for $\nu(r)$, we find:
\begin{equation}\label{nu0}
\nu_0=-\lambda_0=\ln\left(1-\frac{2m}{r}-\frac{\tilde{\Lambda} r^2}{3}\right),
\end{equation}
where $m$ is the constant of integration.

Thus, in the zero approximation, as in the case of the scalar monopole \cite{Yu_TMF24}, we obtain the well-known Schwarzschild-de Sitter solution \cite{Edd} with the cosmological constant renormalized according to \eqref{Lambda_tilde}:
\begin{eqnarray}\label{Shvarc-deSit}
ds^2= \displaystyle \left(1-\frac{2m}{r}-\frac{\tilde{\Lambda} r^2}{3}\right)dt^2  -\left(1-\frac{2m}{r}-\frac{\tilde{\Lambda} r^2}{3}\right)^{-1}dr^2-r^2d\Omega^2.
\end{eqnarray}

It turns out that in the first approximation of the smallness of $\phi$ the solution \eqref{Shvarc-deSit} remains valid. Indeed, due to \eqref{Eq_A}, in the first approximation the relation \eqref{lambda=-nu} is preserved, and therefore the equation \eqref{Eq_B0} is also preserved. Thus, the \emph{ metric \eqref{Shvarc-deSit} is preserved in the approximation linear in $\phi,\varphi$}. Therefore, in the linear approximation, the field equation \eqref{Eq_C} can be considered against the background of the Schwarzschild - de Sitter solution \eqref{Shvarc-deSit}. So, in the linear approximation \eqref{F,f->stabil} we obtain equations for perturbations of the components of the scalar doublet $\{\phi(r),\varphi(r)\}$
\begin{eqnarray}\label{Eq_C20}
\displaystyle \frac{d^2\phi}{dr^2}+\frac{d}{dr}\ln\bigl(r^2\mathrm{e}^{\nu_0(r)}\bigr)\frac{d\phi}{dr}+2m^2_s\phi=0;\\
\label{Eq_F20}
\displaystyle \frac{d^2\varphi}{dr^2}+\frac{d}{dr}\ln\bigl(r^2\mathrm{e}^{\nu_0(r)}\bigr)\frac{d\varphi}{dr}+\mu^2_s\varphi=0.
\end{eqnarray}
By introducing the dimensionless variable $x$ and dimensionless non-negative parameters $\gamma,\sigma,\varrho$
\begin{eqnarray}\label{new_var}
x= \frac{r}{2m};\; \gamma=\frac{4}{3}\tilde{\Lambda} m^2\geqslant0;\; \sigma=2\sqrt{2}mm_s\geqslant0;\; \varrho=\sqrt{2}m\mu_s\geqslant0,
\end{eqnarray}
Let's rewrite the equations \eqref{Eq_C20} -- \eqref{Eq_F20} in terms of these quantities, in which both equations take exactly the same form:
\begin{eqnarray}\label{Eq_C2}
\displaystyle \frac{d^2\phi}{dx^2}+\frac{1-2x+4\gamma x^3}{x(1-x+\gamma x^3)}\frac{d\phi}{dx}+\sigma^2\phi=0;& \\
\label{Eq_F2}
\displaystyle \frac{d^2\varphi}{dx^2}+\frac{1-2x+4\gamma x^3}{x(1-x+\gamma x^3)}\frac{d\varphi}{dx}+\varrho^2\phi=0. &
\end{eqnarray}

Note that coincidence, up to redesignation of the field equations for linear perturbations of the canonical and phantom fields, is achieved precisely near the R-stable point of the corresponding dynamical system.
The stability of solutions leads to their damped oscillatory regime near a stable singular point. Let us also write the equations \eqref{Eq_C2} -- \eqref{Eq_F2} in the form of a normal system of first-order equations:
\begin{eqnarray}\label{Eq_SysC}
\frac{d\phi}{dx}=Z(x);\quad \displaystyle \frac{dZ}{dx}=-\frac{1-2x+4\gamma x^3}{x(1-x+\gamma x^3)} z-\sigma^2\phi;
\label{Eq_SysF}
\frac{d\varphi}{dx}=z(x);\quad \displaystyle \frac{dz}{dx}=-\frac{1-2x+4\gamma x^3}{x(1-x+\gamma x^3)} z-\varrho^2\varphi.
\end{eqnarray}

\subsection{Behavior of solutions near the singularity $r=0$}

When
\begin{equation}\label{gamma>4/27}
\gamma>\frac{4}{27},
\end{equation}
when there are no horizons in the Schwarzschild-de Sitter solution \eqref{Shvarc-deSit}, the solutions to the equations \eqref{Eq_C2} -- \eqref{Eq_F2} are analytic functions on the entire real half-axis $x\in[0,+\infty)$ .
Note, however, that in this case, on the entire semi-axis $x\in[0,+\infty)$
\begin{equation}\label{exp(nu)<0}
\mathrm{e}^{\nu(r)}<0,
\end{equation}
i.e., the entire space is a $\mathrm{T}$-domain, therefore, for the correct interpretation of solutions it is necessary to swap the time and radial coordinates in the Schwarzschild - de Sitter metric \eqref{Shvarc-deSit} \cite{Novikov}. In this case, among other things, the components of the energy-momentum tensor $T^1_1$ and $T^4_4$, and, therefore, the scalars $\varepsilon$ and $p_\parallel$ will also change places.

At $x\to0$ the field equations \eqref{Eq_C2} -- \eqref{Eq_F2} are reduced to simple second order differential equations
\begin{eqnarray}
\phi'' +\frac{\phi'}{x}+\sigma^2\phi=0,& x\to0, & \varphi'' +\frac{\varphi'}{x}+\varrho^ 2\varphi=0,
\end{eqnarray}
who have their decisions
\begin{eqnarray}\label{phi(0)}
\phi(x)= C_1 \mathrm{I}_0(\sigma x)+C_2\mathrm{Y}_0(\sigma x)\backsimeq C_1+C_2\frac{2}{\pi}\ln \sigma x,& (\sigma x \to 0),\\
\label{varphi(0)}
\varphi(x)= \tilde{C}_1 \mathrm{I}_0(\varrho x)+\tilde{C}_2\mathrm{Y}_0(\varrho x)\backsimeq \tilde{C}_1+\tilde{C}_2\frac{2}{\pi}\ln \varrho x,& (\varrho x \to 0)
\end{eqnarray}
where $\mathrm{I}_0(z)$ and $\mathrm{Y}_0(z)$ are Bessel functions of the 1st and 2nd kind, respectively. Thus, the potentials of the scalar field diverge logarithmically near the singularity, and their derivatives are equal
\begin{eqnarray}\label{phi'(0)}
\left.\Phi'\right|_{x\to0}\backsimeq \frac{C_2}{x} \Rightarrow  \left.\frac{d\Phi}{dr}\right|_{r\to0}=\frac{Q}{r}\Rightarrow C_2=\frac{\pi Q}{4m};& \displaystyle
\label{varphi'(0)}
\left.\varphi'\right|_{x\to0}\backsimeq \frac{\tilde{C}_2}{x} \Rightarrow \left.\frac{d\varphi}{dr}\right|_{r\to0}-\frac{q}{r}\Rightarrow \tilde{C}_2=-\frac{\pi q}{4m}.
\end{eqnarray}

Thus, returning to the variable $r$ according to \eqref{new_var}, we obtain for the main terms of the functions $\phi(r)$ and $\varphi(r)$
\begin{equation}\label{phi,varphi_r->0}
\left.\phi(r)\right|_{r\to0}\approx \frac{Q}{2m}\ln\bigl(\sqrt{2}m_sr\bigr);\qquad
\left.\varphi(r)\right|_{r\to0}\approx -\frac{q}{2m}\ln\biggl(\frac{\mu_sr}{\sqrt{2}}\biggr).
\end{equation}
\subsubsection{Behavior of solutions at infinity $\gamma x^3\gg 1$}
In area
\begin{equation}\label{x->8}
x\to\infty,\qquad \gamma x^2\gg 1
\end{equation}
equations \eqref{Eq_C2}, \eqref{Eq_F2} take the form
\begin{equation}\label{EQ_x->8}
\frac{d^2\phi}{dx^2}+\frac{4}{x}\frac{d\phi}{dx}+\sigma^2\psi=0,\quad \frac{d^ 2\varphi}{dx^2}+\frac{4}{x}\frac{d\varphi}{dx}+\varrho^2\varphi=0
\end{equation}
and have their own decisions
\begin{eqnarray}\label{sol_x->8}
\Phi(x)= \pm\frac{m_s}{\sqrt{\alpha}}+\frac{C_1}{x^3}(\sigma x\cos{\sigma x}-\sin\sigma x)+\frac{C_2}{x^3}(\cos{\sigma x}+\sigma x \sin\sigma x),\nonumber\\
\varphi(x)= \frac{C_1}{x^3}(\sigma x\cos{\sigma x}-\sin\sigma x)+\frac{C_2}{x^3}(\cos{\sigma x}+\sigma x \sin\sigma x).
\end{eqnarray}
In this case, we obtain damped oscillations with periods
\begin{equation}\label{tau}
\tau_m=\frac{2\pi}{\sigma}\Rightarrow T_m=\frac{\sqrt{2}\pi}{m_s}; \quad \tau_\mu=\frac{2\pi}{\varrho}\Rightarrow T_\mu=\frac{2\sqrt{2}\pi}{\mu_s},
\end{equation}
whose amplitude decreases in inverse proportion to $x^2$.

Note that the damping of oscillations of perturbations of an asymmetric scalar doublet is precisely evidence of its stability in the state
\eqref{x8}
\begin{equation}\label{Stabil_Dublet}
\mathbf{S_0}:\ \left\{\Phi_0=\pm \frac{m_s}{\sqrt{\alpha}},\varphi_0=0\right\}.
\end{equation}

In Fig. \ref{Ignatev1} and \ref{Ignatev2} show different parts of the graph of solutions to the equation (on different scales) \eqref{Eq_C2} (or \eqref{Eq_F2}), obtained by numerical integration of this equation for parameter values
\begin{equation}\label{par}
\mathbf{P=[\gamma,\sigma,\varrho]}=[0.2,\sigma,\varrho]
\end{equation}
and initial conditions
\begin{equation}\label{Ini}
I=[x_0,\phi_0=\phi(x_0),Z_0=Z(x_0),\varphi_0=\varphi(x_0),z_0=z(x_0)]=[1,1,0,1,0].
\end{equation}
In this case, solid lines indicate graphs corresponding to the parameters: $\sigma,\varrho=1$, dashed: $\sigma,\varrho=0.5$, dash-dotted: $\sigma,\varrho=2$. As is easy to see, for large values of $x$ these graphs describe damped oscillations with a period \eqref{tau} inversely proportional to $\sigma,\varrho$.

\TwoFigss{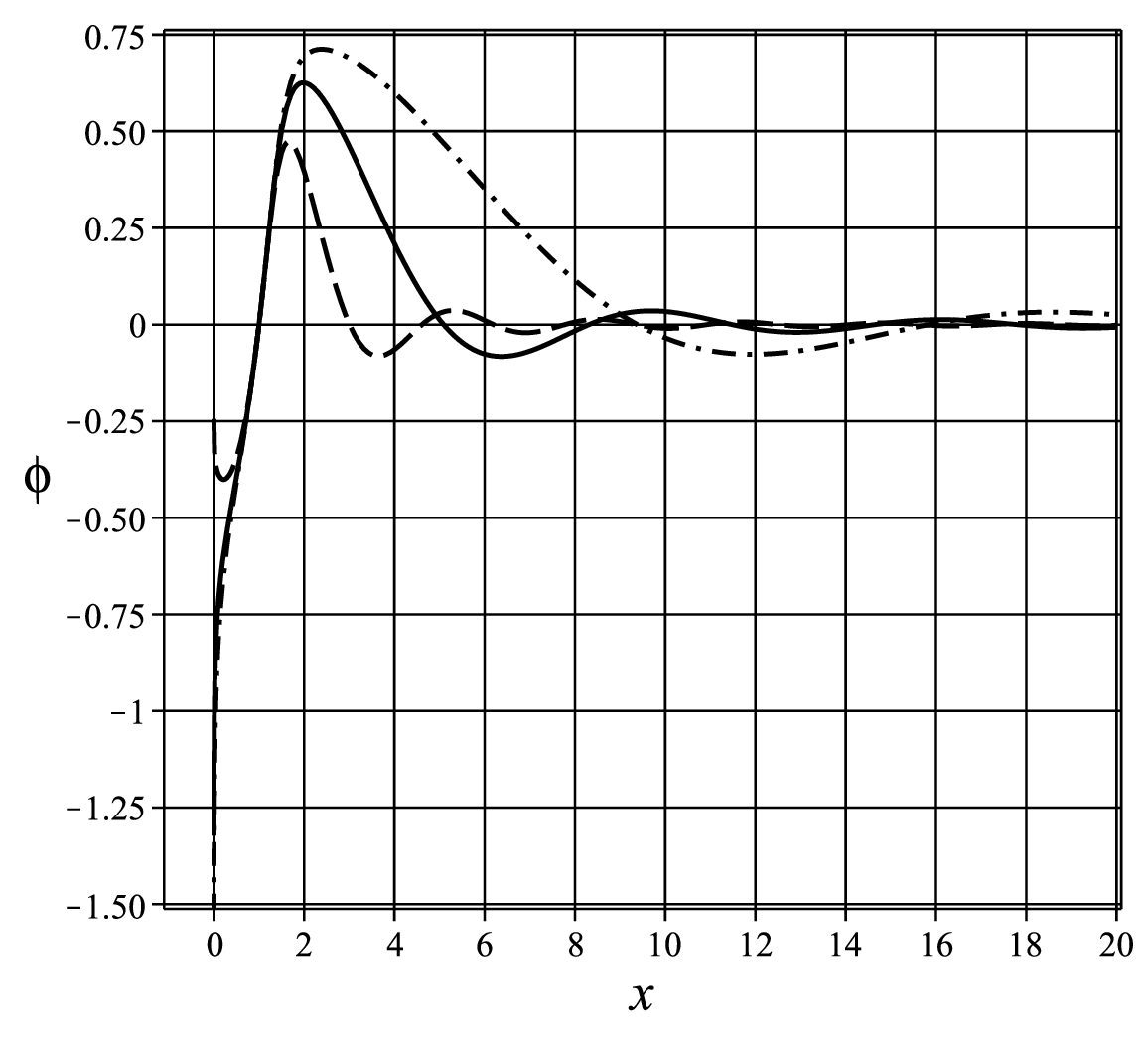}{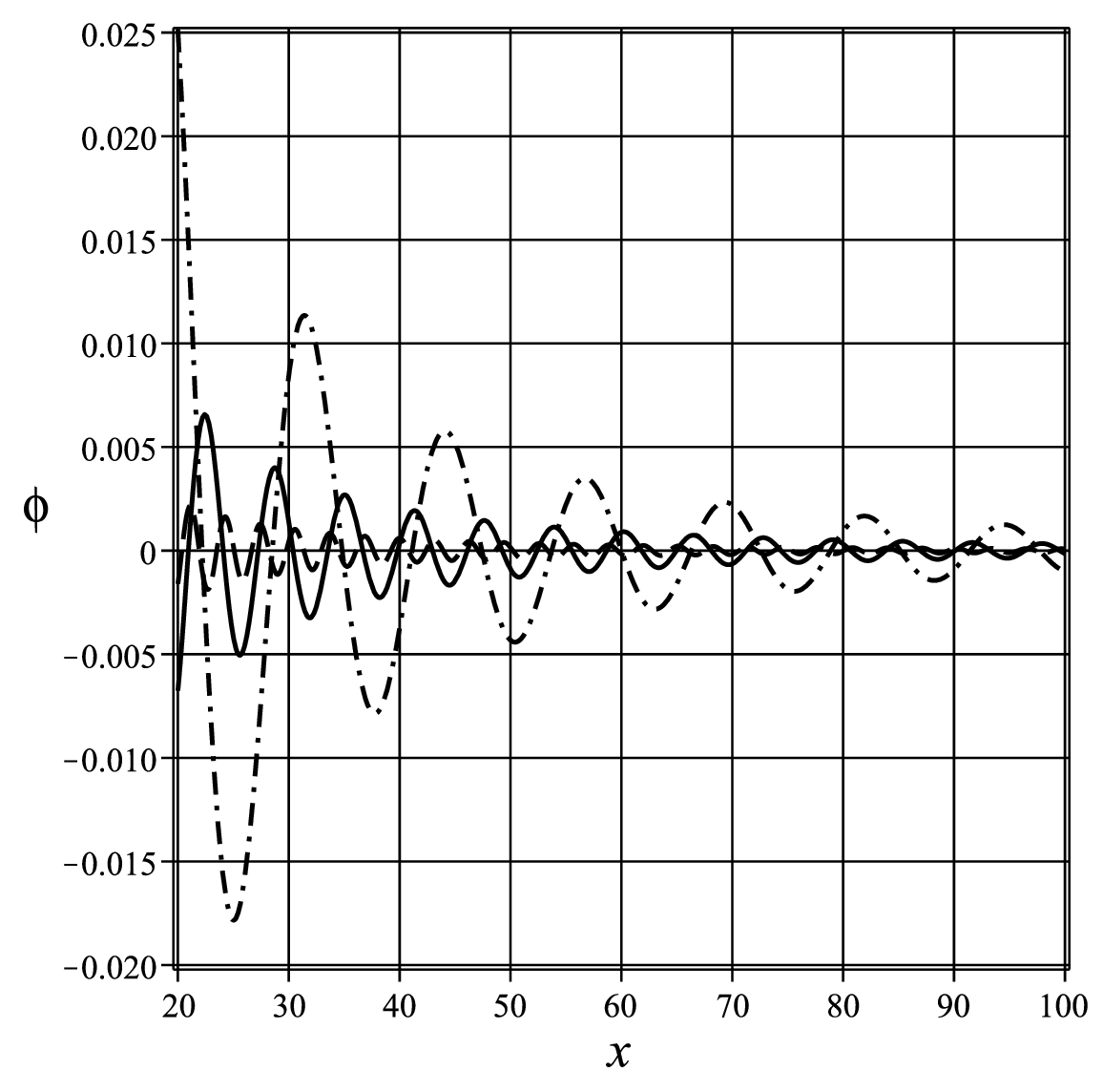}{\label{Ignatev1}Graphs of potential perturbation $\phi(x)$ (or $\varphi(x)$) for the values of parameters \eqref{par} and initial conditions \eqref{Ini} on the segment $x\in[0,20]$.} {\label{Ignatev2}Graphs of potential perturbation $\phi(x)$ (or $\varphi(x)$) for the values of parameters \eqref{par} and initial conditions \eqref{Ini} on the interval $x\in[20,100 ]$.}

In the case of small values of $\gamma$, an intermediate region can form with oscillations of the scalar field damping in proportion to $1/x$, which then quickly fall (\cite{Yu_TMF24}):
\begin{eqnarray}\label{1/x}
\gamma\ll 1,\qquad x\in \left(1,\frac{1}{\sqrt{\gamma}}\right):  \displaystyle \phi\backsimeq \frac{\mathrm{e}^{i\sigma x}}{x},\; \varphi\backsimeq \frac{\mathrm{e}^{i\varrho x}}{x} ; \\
\label{1/x^2}
 \forall\gamma, \qquad \displaystyle x\in \left(\mathrm{Max}\left\{\frac{1}{\sqrt{\gamma}},1\right\},+\infty\right):  \displaystyle \phi\backsimeq  \frac{\mathrm{e}^{i\sigma x}}{x^2},\; \varphi\backsimeq \frac{\mathrm{e}^{i\varrho x}}{x^2}.
\end{eqnarray}
\section{Macroscopic averages and oscillation energy}
\subsection{Averaging of oscillations near a state of stable equilibrium}
Oscillations of a scalar doublet are microscopic oscillations \emph{with Compton wavelengths } $\exp(i\sqrt{2}m_s r)$, $\exp(i\mu_s r)$. A macroscopic observer measures only macroscopic characteristics, such as macroscopic energy density and pressure. Taking into account the rapidly oscillating nature of the solutions to the system of equations \eqref{Eq_SysC} --\eqref{Eq_SysF} at
\begin{eqnarray}\label{WKB}
\sigma x\gg1 \Rightarrow m_s r \gg 1, & \varrho x\gg1 \Rightarrow \mu_s r\gg1,
\end{eqnarray}
Let us average the values of \eqref{T^i_k} over a sufficiently large interval of the radial variable
\begin{eqnarray}\label{T}
T\gg \mathrm{max}\{T_m,T_\mu\},\nonumber
\end{eqnarray}
introducing according to \cite{Yu_TMF24} the macroscopic average of the rapidly varying function $f(r)$:
\begin{equation}\label{everage}
\overline{f(r)}= \frac{1}{T}\int\limits_{r-T/2}^{r+T/2}f(r)dr\Rightarrow \overline{f(x)}= \frac{1}{\tau}\int\limits_{x-\tau/2}^{x+\tau/2}f(x)dx.
\end{equation}
Let $f(r)$ be a smooth function, so that $f'(r)\sim f/r$, $kr\gg1$. Then in the WKB approximation \eqref{WKB} the following relations are valid:
\begin{equation}
\overline{f(r)\sin^{2n}(k r)}\backsimeq \overline{f(r)\cos^{2n}(k r)} \approx \frac{(2n-1)!!}{2n!!}f(r);\quad \overline{f(r)\sin^{2n+1}(k r)}\backsimeq \overline{f(r)\cos^{2n+1}(k r)} \approx 0
\end{equation}

Since the solutions we are considering are oscillatory in nature, the macroscopic averages of odd powers of potentials and their derivatives vanish. Thus, expanding the energy-momentum tensor \eqref{T^i_k} in the quadratic approximation in the smallness of $\phi(r),\varphi(r)$ near the state of stable equilibrium $\mathbf{S_0}$ \eqref{Stabil_Dublet}, we get:
\begin{eqnarray}\label{T_44_2}
8\pi T^4_4\equiv 8\pi\varepsilon=\displaystyle -\frac{\mu^4_s}{4\beta}+\frac{\mathrm{e}^{\nu_0(r)}}{2}({\phi'}^2-{\varphi'}^2)-m^2_s\phi^2+\frac{1}{2}\mu^2_s\varphi^2;\\
\label{T_11_2}
\displaystyle -8\pi T^1_1=8\pi p_\parallel=\frac{\mu^4_s}{4\beta}+\frac{\mathrm{e}^{\nu_0(r)}}{2}({\phi'}^2-{\varphi'}^2)+m^2_s\phi^2-\frac{1}{2}\mu^2_s\varphi^2.
\end{eqnarray}

As can be seen from the previous, in all cases the following asymptotic solutions of field equations are valid (either \eqref{Phi,varphi=0+e} for flat space, or \eqref{1/x} -- \eqref{1/x^2 } for the Schwarzschild-de-Sitter field:\footnote{More precisely, instead of $\cos$ you need to write a linear combination of $\cos$ and $\sin$, but this clarification does not affect\\ the final result.}
\begin{eqnarray}\label{phi-sim}
\phi(r)\backsimeq \cos(\sqrt{2}m_s r)\frac{\phi_0}{r^n}; & \displaystyle \phi'(r)\backsimeq  -\sin(\sqrt{2}m_s r)\frac{\sqrt{2}m_s\phi_0}{r^n};&\nonumber\\
\varphi(r)\backsimeq \cos(\mu_s r)\frac{\varphi_0}{r^n}; & \displaystyle \varphi'(r)\backsimeq  -\sin(\mu_s r)\frac{\mu_s\varphi_0}{r^n};& (n=1,2),
\end{eqnarray}
we get
\begin{eqnarray}\label{asimpt_Ff}
\overline{\phi(r)}\approx 0; & \overline{\phi'(r)}\approx 0; & \overline{\varphi(r)}\approx 0; \hskip 1.1cm \overline{\varphi'(r)}\approx 0;\\
\overline{\phi^2(r)}\approx \frac{|\phi_0|^2}{2r^{2n}}; & \displaystyle \overline{\phi'^2(r)}\approx \frac{m^2_s|\phi_0|^2}{r^{2n}}; &
\overline{\varphi^2(r)}\approx \frac{|\varphi_0|^2}{2r^{2n}}; \hskip 0.4cm \displaystyle \overline{\varphi'^2(r)}\approx \frac{\mu^2_s|\varphi_0|^2}{2r^{2n}}.\nonumber
\end{eqnarray}
Renormalizing the cosmological constant according to \eqref{Lambda_tilde}, which corresponds to the removal of the constant term in \eqref{T_44_2} - \eqref{T_11_2} and substituting \eqref{asimpt_Ff} into the resulting expressions, we obtain for the macroscopic average energy density and pressure of oscillations of the scalar doublet :
\begin{eqnarray}\label{p,e_psi_average}
\displaystyle 8\pi\overline{\varepsilon}=(2m_s^2|\phi_0|^2-\mu_s^2|\varphi_0^2|)\frac{\mathrm{e}^{\nu_0(r)}-1}{4r^{2n}}; & \displaystyle 8\pi \overline{p_\parallel}=(2m_s^2|\phi_0|^2-\mu_s^2|\varphi_0|^2)\frac{\mathrm{e}^{\nu_0(r)}+1}{4r^{2n}}.
\end{eqnarray}

Note, firstly, that the left-hand sides of the relations \eqref{p,e_psi_average} are expressions for the dimensionless normalized macroscopic average energy density and radial pressure fields of the scalar doublet.

Secondly, note that in flat space-time $\nu(r)=0\Rightarrow \mathrm{e}^{\nu_0(r)}-1=0$, $\mathrm{e}^{\ nu_0(r)}+1=2$. Therefore, in the absence of a gravitational field, we obtain for macroscopic energy densities
and pressure of the scalar field (taking into account the renormalization of the cosmological constant.)
\begin{eqnarray}\label{p,e_psi_average,M=0}
\displaystyle \mathbf{M=0}:\quad 8\pi\overline{\varepsilon}=0; & \displaystyle 8\pi \overline{p_\parallel}=\frac{2m_s^2|\phi_0|^2-\mu_s^2|\varphi_0|^2}{2r^2}
\end{eqnarray}
-- the macroscopic energy density of scalar fields is zero.

Thirdly, we note that, in contrast to the case of a classical scalar singlet \cite{Yu_TMF24}, the macroscopic oscillation energy density is no longer necessarily negative. Really,
\begin{eqnarray}\label{exp(nu)+-1}
\mathrm{e}^{\nu_0(r)}-1=-\left(\frac{2M}{r}+\frac{\tilde{\Lambda}}{3}r^2\right)<0; & \displaystyle \mathrm{e}^{\nu_0(r)}+1=\left(2-\frac{2M}{r}-\frac{\tilde{\Lambda}}{3}r^2\right).
\end{eqnarray}
Thus, we obtain for $M\not=0$:
\begin{eqnarray}\label{p,e_psi_average,Mnot=0}
\displaystyle 8\pi\overline{\varepsilon}=-\left(\frac{M}{r}+\frac{\tilde{\Lambda}}{6}r^2\right)\frac{2m_s^2|\phi_0|^2-\mu_s^2|\varphi_0|^2}{2r^{2n}}; & \displaystyle 8\pi \overline{p_\parallel}=\left(1-\frac{M}{r}-\frac{\tilde{\Lambda}}{6}r^2\right)\frac{2m_s^2|\phi_0|^2-\mu_s^2|\varphi_0|^2}{2r^{2n}},
\end{eqnarray}
so for a purely classical scalar field ($\varphi=0$) $\overline{\varepsilon}<0$ (see \cite{Yu_TMF24}), for a purely phantom field $\phi=0$, on the contrary, $\ overline{\varepsilon}>0$.

In the region $\forall \gamma$ according to \eqref{1/x^2} we obtain from \eqref{p,e_psi_average,Mnot=0}
\begin{equation}\label{e_r->8}
\left.\overline{\varepsilon}\ \right|_{\gamma x^3\to\infty}=\left.\overline{p_\parallel}\ \right|_{\gamma x^3\to\infty} \backsimeq -\frac{\tilde{\Lambda}}{96\pi r^2}(2m_s^2|\phi_0|^2-\mu_s^2|\varphi_0|^2).
\end{equation}

Note that in this far region
\begin{equation}
\bar{p}_\parallel\approx \bar{p}_\perp=\bar{\varepsilon}, \qquad (\forall \gamma)
\end{equation}
-- the macroscopic medium is isotropized and described by an extremely rigid equation of state.

\subsection{Orientation of the scalar doublet}
In what follows, under the condition $2m_s^2|\phi_0|^2-\mu_s^2|\varphi_0|^2>0$, we will call the scalar doublet \emph{canonically oriented}, under the condition $2m_s^2|\phi_0|^ 2-\mu_s^2|\varphi_0|^2<0$ -- \emph{phantom-oriented}, and in the case
\begin{equation}\label{e=0}
2m_s^2|\phi_0|^2-\mu_s^2|\varphi_0|^2=0
\end{equation}
-- \emph{neutral}. From \eqref{p,e_psi_average,M=0}, \eqref{p,e_psi_average,Mnot=0} and \eqref{e_r->8} it follows that in all cases for a neutral scalar doublet $\overline{\varepsilon} =\overline{p_\parallel}=0$ -- \emph{macroscopic energy and pressure densities are strictly zero}. In this regard, let us find out what microscopic parameters a neutral scalar doublet corresponds to.

From \eqref{phi(0)} -- \eqref{varphi'(0)} for $r\to0$ we find:
\begin{equation}\label{F,f_r->0}
\phi\backsimeq \frac{Q}{2m}\ln\bigl(\sqrt{2}m_s r\bigr);\qquad \varphi \backsimeq -\frac{q}{2m}\ln\biggl(\frac {\mu_s}{\sqrt{2}} r\biggr).
\end{equation}
From \eqref{F,f_r->0} it follows
\begin{equation}\label{F/f=const}
\mu_s=2m_s \Rightarrow \left.\frac{\phi(r)}{\varphi(r)}\right|_{r\to0}=\frac{Q}{q}=\mathrm{Const}.
\end{equation}
Using this value in the definition of a neutral scalar doublet \eqref{e=0}, we obtain $Q^2=2q^2$, i.e., the following fundamental parameters correspond to a neutral scalar doublet:
\begin{equation}\label{neutral}
\mu_s=2m_s;\qquad q=\pm \frac{1}{\sqrt{2}} Q.
\end{equation}

Rewriting now the neutrality condition for the doublet \eqref{e=0} in terms of fundamental parameters, we obtain the condition for its neutrality:
\begin{equation}\label{neutral1}
\Upsilon\equiv 2m^2_s Q^2-\mu^2_s q^2=0.
\end{equation}
In the case of a positive left-hand side \eqref{neutral1} $\Upsilon>0$ the scalar doublet will be canonically oriented, otherwise it will be phantom oriented.

\subsection{Mass of the scalar halo of a black hole}
As we already noted in the first part of the article \cite{Yu_TMF24}, microscopic oscillations of the scalar field are perceived by an external macroscopic observer as some liquid with macroscopic energy density and anisotropic pressure $\bar{p}_\parallel\not=\bar{p}_ \perp=\bar{\varepsilon}$. This liquid can be called the \emph{scalar halo of a black hole}. A scalar halo can have its own mass (see, for example, \cite{Land_Field})
\begin{eqnarray}
M_s(r)=4\pi\int\limits_0^r T^4_4 r^2dr.
\end{eqnarray}
Since microscopic oscillations occur in the far region $r\to\infty$, the mass of the scalar black hole halo can be estimated using \eqref{e_r->8}:
\begin{eqnarray}\label{M_s}
M_s(r)=4\pi \int\limits_{r_0}^r \overline{\varepsilon} r^2dr \sim - \frac{\tilde{\Lambda}}{24}(2m^2_s|\phi_0| ^2-\mu^2_s|\varphi_0|^2)(r-r_0),
\end{eqnarray}
where $r_0\sim \max(m_s,\mu_s)$.

Thus, in the case of a canonically oriented scalar doublet, the mass of the scalar halo is negative and reduces the observed mass of the black hole $m$; in the case of a phantom oriented scalar doublet, the observed mass grows and, in principle, can tend to infinity. In the case of a neutral scalar doublet, the observed black hole mass is conserved.

\section*{Conclusion}
Summing up the article, we note its main results.
\begin{itemize}
\item A mathematical model of a self-gravitating asymmetric Higgs scalar doublet in the case of spherical symmetry has been constructed and studied.
\item It is shown that $R$-stable solutions correspond to solutions close to stable points of the Higgs potential \eqref{stab_F,f} with negative total energy of the scalar doublet
\[\varepsilon_{min}=-\frac{\mu_s}{32\pi\beta}.\]
Moreover, in the far region, solutions to field equations for scalar fields have the character of damped microscopic oscillations around a stable solution.
\item It is shown that stable solutions correspond to the Schwarzschild - de Sitter metric with the value of the cosmological constant reformatted by the negative energy of the scalar doublet.
\item It is shown that in the far zone $r\to\infty$ the macroscopic medium is isotropic and corresponds to an ideal fluid with an extremely rigid equation of state.
\item The macroscopic average energy densities and pressures of scalar oscillations are calculated and it is shown that the total macroscopic energy density of oscillations is determined using the fundamental constants of the scalar
charge of some constant $\Upsilon$ \eqref{neutral1}: for $\Upsilon>0$ the energy density is negative (canonically oriented doublet), for $\Upsilon<0$ the energy density is positive (phantom oriented doublet),
at $\Upsilon=0$ the energy density is zero (neutral doublet).
\item It is shown that a neutral scalar doublet corresponds to certain relationships between its fundamental constants.
\end{itemize}

Let us note, firstly, that in the case of a phantom oriented scalar doublet, the growth of the mass of the scalar halo as $r\to\infty$ is, in principle, unlimited, which casts doubt on the possibility of the existence of such a doublet.
Secondly, note that in the case of a canonically oriented scalar doublet, the increase in the negative mass of the scalar halo is limited by the mass of the black hole. In the limiting case $M_s=-m$, the total mass of the black hole disappears. Therefore, further growth of negative energy, apparently, should stop at some $r=r_\infty$. In this case, an infinitely distant observer will record the total zero mass of the object, although in the vicinity of the black hole the gravitational field will be strong. According to \eqref{M_s}, the halo radius of this object is
\[r_\infty\backsimeq \frac{24m}{\tilde{\Lambda}(2m^2_s|\phi_0|^2-\mu^2_s|\varphi_0|^2)}.\]
Finally, thirdly, we note that in the case of a neutral scalar doublet, the macroscopic energy density, pressure and mass of the halo are strictly equal to zero, i.e., from the point of view of a macroscopic observer, the scalar halo is completely absent. This fact, as well as the stability of the corresponding field solutions, allows us to formulate an assumption about the possible existence of a superheavy heterogeneously scalarly charged particle with the fundamental parameters of the Higgs potential \eqref{neutral}.
\subsection*{Funding}
The work was carried out in accordance with the Program of the Government of the Russian Federation for strategic academic leadership ``Priority 2030'' of the Kazan Federal University.

\setcounter{section}{0}
\setcounter{equation}{0}
\setcounter{figure}{0}



\begin{thebibliography}{75}
\bibitem{Yu_TMF24}
Yu.G. Ignat'ev, ``Self-gravitating Higgs field of a scalar charge'', Theor. Mat. Phys., \textbf{219}, (2024) (to be published).

%
\bibitem{YuKokh_TMF}
 Yu. G. Ignat'ev,  I. A. Kokh, ``Complete cosmological model based on an asymmetric scalar Higgs doublet'', Theor. Mat. Phys., \textbf{207:1}, 514-552  - (2021).
%
\bibitem{Land_Field}
L. D. Landau, E. M. Lifshitz. \emph{The Classical Theory of Fields.} Pergamon Press. Oxford$\cdot$ New York$\cdot$ Toronto$\cdot$ Sydney$\cdot$ Paris$\cdot$ Frankfurt, 1971.
%

%
\bibitem{Bogoyav}
 O. I. Bogoyavlensky, The methods of the qualitative theory of dynamic systems in astrophysics and gas dynamics. Moskow, Nauka (1980).
%

\bibitem{Scalar_Charge}
Yu. G. Ignat'ev, ``Scalarly Charged Particles and Particle Interaction with the Higgs Potential''	\emph{Gravit. Cosmol.}, \textbf{29:3}, 213-219, (2023).
%

%
\bibitem{Novikov}
Ya.B. Zel'dovich, I.D. Novikov, \emph{The theory of gravity and the evolution of stars}, Moskow: Nauka (1971) (in Russian);
Ya.B. Zel'dovich, I.D. Novikov, \emph{Stars and Relativity}, Dover Publ. Inc., New York (1971) (trnslate).
%

\bibitem{Edd}
A. S. Eddington, \emph{Mathematical Theory of Relativity}, Cambridge Univ. Press, Cambridge (1923).


\end{thebibliography}
\end{document}